\documentclass[12pt]{article}
\usepackage{amsmath}
\usepackage{graphicx}
\usepackage{wrapfig} 
\usepackage{url}

\DeclareMathOperator{\Namei}{\mathcal{N}}

\DeclareMathOperator{\unixfs}{unix\!f\!s}
\DeclareMathOperator{\nulls}{Null}

\DeclareMathOperator{\aappend}{\mathbf{\backslash}}

\DeclareMathOperator{\initial}{\iota}
\newcommand{\fwrite}[1]{\mathtt{write}[#1]}
\newcommand{\fdelete}[1]{\mathtt{delete}[#1]}

\begin{document}
\title{Folding a tree into a map}

\author{Victor Yodaiken}
\maketitle
\begin{abstract}
Analysis  of the retrieval architecture of the highly influential UNIX file system  (\cite{Ritchie}\cite{multicsfs}) provides insight into design methods, constraints, and possible alternatives. The basic architecture  can be understood  in terms of function composition and recursion by anyone with some mathematical maturity. Expertise in operating system coding or in any specialized ``formal method'' is not required. 
\end{abstract}

\section{Basics}
\begin{wrapfigure}{r}{0.38\textwidth}
\includegraphics[width=0.30\textwidth]{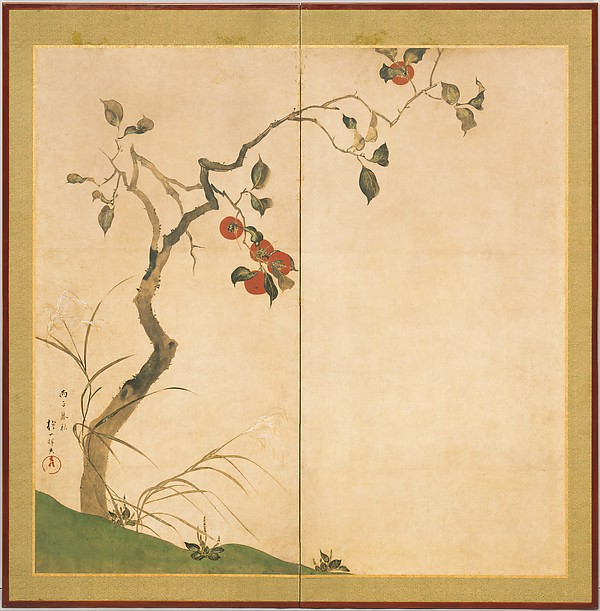}
\cite{persimon}
\end{wrapfigure}
The retrieval (read-only) operation of computer file system can be represented by a map: \[File:Identifiers \to Contents.\] A disk drive corresponds to a much simpler map \[Disk:BlockNumbers \to Blocks\] where blocks are just fixed length sequences of  ``bytes''  ( 8 adjacent binary digits). Until recently, file systems had to be designed around the problem of building a reasonably sophisticated map of the first kind from the simple operation of second map. The  UNIX file system  (\cite{Ritchie}\cite{multicsfs})  supports variable sized files from small text files to videos and databases and also organizes files into a tree structure so that file names describe paths through the tree.  The map looks like

\[ F:Paths(X) \to Contents.\] 
The set $Paths(X)$ is the set of  finite sequences of non-null strings over some alphabet $X$.  Implementations use special characters as separators --- e.g. $/a/b/c$ where  $/$  but fundamentally paths are sequences of strings\footnote{ In practice, the length of paths and lengths of strings in the path will be constrained by some bound, but we don't have to worry about that now either.}.
 
  The UNIX designers split the problem of building this system on top of a disk drive into two conceptually distinct problems. First they looked at how to get past the fixed size:
  \[\alpha: Indexes\to Contents\]
  where $Indexes$ is a set of numbers and $Contents$ includes, at least, all sequences of bytes within the limits of the file system. 
 The second step is to embed the tree into this first system:
 \[\beta: Paths(X)\to Indexes\]
 The file system map is then given by: $F(p) = \alpha(\beta(p))$.  The map $\beta$ relies on a clever technique where $Contents$  is the union of the set of byte sequences (ordinary files) and the set of   maps $Strings(X)\to Indexes$ (\emph{directories}).  If $\alpha(i)$ is a directory, then  $\alpha(i)(x)$ applies the map that is the value of $\alpha(i)$ to the argument $x$.  
 
 A recursive descent from an initial index  processes strings starting on the left. Let $\nulls$ denote the null path (0 strings).  If $p$ is a path and $x$ is a non-null string, then $xp$ is the path  obtained by appending $x$ to $p$ on the left. Then every finite path is either the null path or of the form $xp$.
 \[\begin{array}{l}\Namei( i,\nulls) = i\\
\Namei(i, xp) = \Namei( \alpha(i)(x),p)\end{array}\]
At each step, $\Namei$ resolves the leftmost string in the path -- assuming that $\alpha(i)$ is  a directory that is defined on that string.

To define $\beta$ then we just need to pick some $root\in Indexes$ and then:
\[\beta(p) = \Namei(root,p)\]
\section{Consistency and extending the model}
One useful property of $\Namei$ is that it is guaranteed to terminate:
\begin{equation}
\mbox{If }p\mbox{ is }n\mbox{ elements long }\Namei(i,p)\mbox{ terminates in at most }n+1\mbox{ steps}\label{eq:terminates}
\end{equation}
 This property, assures implementors that their program will not cycle infinitely trying to reduce a path to an index. A second property  implicit in the definition of $\Namei$ is that if a path names a file, every prefix of that path names a directory. Let $px$ be the path obtained by appending $x\in Strings(X)$ to $p\in Paths(X)$ on the right. 
 \begin{equation}
\mbox{If }F(px)\mbox{ is defined  then }F(p)\mbox{ is a directory}\label{eq:prefix}
\end{equation}
There are a number of additional properties that $\alpha$ and $root$ need to satisfy to make this a reasonable file system. The most important are the ``no orphans'' and ``no dangling references'' properties. 
 \begin{eqnarray}
  \mbox{ If }\alpha(i)\mbox{ is defined, there is some path }p\mbox{ s.t. } \Namei(root,p)=i\label{orphan}\\
F(px)\mbox{ is defined if and only if }F(p)(x)
  \mbox{ is defined.}\label{dangle}
  \end{eqnarray} 
  
There's another useful pair of properties for two special strings $''.''$ (self) and $''..''$ (parent).
   \begin{eqnarray}
 \mbox{if }F(p)\mbox{ is a directory, then } F(p)(''.'')=\Namei(root,p)\\
\mbox{if } F(px)\mbox{ is a directory, then }F(px)(''..'') = \Namei(root,p)\label{eq:parent}\\
F(root)(''..'')=root \label{eq:rootparent}
\end{eqnarray}

Do we want to permit aliases - distinct paths that don't contain any ``..''  or ``.''strings but that name the same file?  Say a path is ``dot free'' if it doesn't contain either ``..'' or ``.'' as elements (other dots are ok). 
We can require that for any two dot-free paths  $p,q$:
if $\Namei(root,p)=\Namei(root,q)$  then $p=q$.  The most important aspect of a property like this is the elimination of loops. Consider a program that finds all the files that are rooted by a particular folder.
Let $list(F(p))=\emptyset$ if $p$ is not defined or is not a directory and the set of all strings $x\notin\{``.'',''..''\}$ where $F(p)(x)$ is defined and a directory. Then:
\[
find(p) =\begin{cases}
\emptyset &\mbox{if }F(p) \mbox{ is not defined}\\
\{p\} & \mbox{if } F(p)\mbox{ is a regular file}\\
&\mbox{or } F(p)\mbox{ is the empty directory}\\
\{p\}\cup (\cup_{x\in list(F(p))} find(px))&\mbox{otherwise}\end{cases}\]
The  question is whether $find$ is guaranteed to terminate. Even if $F$ describes a file system with a finite number of files (always the case in practice), loops would cause $find$ to build longer and longer paths without ever hitting one of the first two terminating cases.  If there are at most $n$ files in the system, then a path of more than $n$ elements must visit the same directory twice --- implying there is an alias. So prohibiting aliases is sufficient to make $find$ and many other related algorithms work properly.  The original UNIX file system did not prohibit all aliases, but had a weaker constraint that is enough to assure that $find$ terminates. Define $links$ as follows:
\[ links(i,j) =\begin{cases}
1&\mbox{if }\alpha(i)\mbox{is a directory}\\
&\mbox{and  there is some }x\in list(\alpha(i))\mbox{ s. t. }\alpha(i)(x)=j\\
0&\mbox{otherwise}\end{cases}\]
The requirement is that $\Sigma_{j\in Indexes}links(j,i) < 2$ for all $i$ where $\alpha(i)$ is a directory plus a requirement that $links(j,root) =1$ only if $j=root$. That is,  at most one directory $j$ links to directory $i$ and none link to the root directory.  This constraint is a consequence of the consistency properties above. Suppose that $j_1 \neq j_2$ violated the constraint - so that $\alpha(j_1)(x) = \alpha(j_2)(y)=i$ and $\alpha(i)$ is a directory. Because of property \ref{dangle} there must be $p$ and $q$ so that $\Namei(root,p)=j_1$ and $\Namei(root,q)=j_2$. But   then $F(px)(''.'')=F(qy)(''.'')=i$ and $\alpha(i)(x)=F(px)(''..'')=F(qy)(''..'')$ so  $j_1=j_2$ which contradicts the premise. If $links(i,root)$ then the same reasoning tells us $i=root$ by \ref{eq:rootparent}.

In later versions of UNIX things became more complex because of so called ``soft links'' (symbolic links).  To include soft-links in the current analysis, add a third file type to ordinary files and directories: a soft link type where the contents is just a path. Then modify $\Namei$ as follows:
\[\Namei(i, xp) = \begin{cases}
                                                     \Namei( (\alpha(i))(x),p)&\mbox{if }\alpha(i)\mbox{ is a directory map}\\
                                                      \Namei(root,Concatenate(\alpha(i),xp))&\mbox{if }\alpha(i)\mbox{ is a soft link}\end{cases}\]
Sadly, this new version of $\Namei$ lacks property $\ref{eq:terminates}$. The normal method for fixing that is to count the number of soft-links visited along a path. 
\[\Namei(i,p) = \Namei'(i,p,1)\]
\[\Namei'(i, xp,n) = \begin{cases}
         \Namei'( (\alpha(i))(x),p,n)&\mbox{if }\alpha(i)\mbox{ is a directory map}\\
                                                      \Namei'(Concatenate(\alpha(i),xp),n-1)&\mbox{if }\alpha(i)\mbox{ is a soft link and }n>0\end{cases}\]
 Soft links, however, introduce loops into the tree and $\Namei$ might visit the same directory twice.
 
We have to use a similar trick to make $find$ safe with some $k>0$ as the number of acceptable soft links to traverse. 

Extensions such as mounted file systems and union file systems are easy to add to this model.
\section{Coding\label{sec:coding}}

A disk block can be considered to be just a fixed length sequence of  ``bytes'' ( 8 binary digits representing numbers in the set $\{0 \dots 255\}$). Ordinary files can be specified on the disk by a number $n$ indicating how many bytes are in the file and a sequence of disk block numbers $b_1\dots b_m$. The file contents is then the result of 
concatenating $Disk(b_1)\dots Disk(b_m)$ and then truncating to get $n$ bytes of data.  If the file is a directory, then this is just the first step and the second step is to decode the directory from the data.  For example, if file names are composed of 16-bit unicode, then the contents of a directory file might be sequences of two byte quantities coding unicode characters, terminated by two 0 bytes, followed by, say, 4 bytes coding an index number.  If ``passwords'' should be mapped to $34832$ then hexadecimal encoding looks something like this:
\[\mathtt{70,61,73,73,77,6F,72,64,73, 00,00, 00,00,88,10}\]
The actual coding is interesting and important, but for this paper, I just want to sketch out one method for concreteness so that
\[ file: Integers \times SequenceOfBlockNumbers \to Contents\]
and 
\[ DecodeDirectory: Contents \to \{Strings\to Indexes\}\]
seem plausible.

The map $\alpha$ depends on similar encoding.  To start, assume we can encode both the length $n$ and the sequence of block numbers of a file in a single block. Let's also encode in that block a ``type'' that tells us if a file is ordinary, directory, or soft link. The disk drives that were the design targets of the UNIX file system could store somewhere around $2^{22}$ bytes of data in $2^{13}$ blocks.  Disk drives at the time of the writing of this paper can easily hold $2^{42}$ or more bytes in $2^{33}$ blocks. In either case, a single disk block cannot hold enough disk block numbers for a really big file so some of the disk block numbers in the sequence encoded in the block are used as indirect numbers - they point to blocks that encode more numbers. The details of this are not covered here - see \cite{bsd} for explanations. 

\[
 \begin{array}{|l|}
 \hline
 \alpha: Indexes\to Contents\\
DecodeSequence:Blocks \to  SequencesOfBlockNumbers\\
DecodeSize:Blocks \to Integers\\
 DecodeType:Blocks \to \{ordinary, directory, softlink\}\\
 \alpha_1: Indexes \to BlockNumbers\\
 \alpha2_(i) = file(DecodeSize(Disk(\alpha_1(i))),DecodeSequence(Disk(\alpha_1(i))))\\
 \alpha(i) = \begin{cases}
 \alpha_2(i)&\mbox{if } DecodeType(Disk(\alpha_1(i)))\in \{ordinary,softlink\}\\
  DecodeDirectory(\alpha_2(i))&\mbox{if } DecodeType(Disk(\alpha_1(i)))=``directory''\\
 undefined&\mbox{otherwise}\end{cases}\\
 file(n,\mathbf{x}) =  truncate(n, Concatenate(Disk(x_1) \dots Disk(x_n) ) )\\
\hline 
 \end{array}
 \]

\section{Discussion \label{sec:discussion}}
The efficiency advantages of the decomposition above are reasonably obvious to anyone with an intuition about system programming but we can also make an informal complexity analysis.  Think of file system maps as finite sets of pairs: in practice file systems are finite. Searching for a file in a map $Paths(X)\to Contents$ would, on average take $n/2$ steps where $n$ is the number of pairs.  This search would involve testing sequences to see if they are equal on each step. This means each step of the search involves multiple steps to compute the match.  To speed up this search, we need to map paths to some sorted data  rather than an unordered set.  That is what the embedding does. Directory maps involve much simpler matching because we are matching strings not sequences and directories should generally be small. In practice, file systems can easily contain billions of files, but directories tend to contain just a few entries. If the average size of a directory is $k$ elements, then  an $n$ element file system will average a depth of only $\log_k(n)$. So for a file system containing one billion ordinary files with average of 10 elements in a directory, 9 steps through the tree would resolve an average path with each step requiring comparison of a string (not a path) to an average of 5 other strings taking us to 45 string matches plus 9 lookups of directories. Compare that to  500,000,000 path matches. 

Consider common queries on the file system such as ``find(p)''. This is efficient to compute using the tree structure.  The UNIX command for a detailed list is also efficient to compute with the embedded tree structure. Detailed list involves extending out the index block(s) to contain additional information - such as last modification and  security/permission data. In this case, $\Namei$ provides an index and $\alpha_1$ provides the block itself. During the 1980s a number of development groups all made the same discovery that detailed list was easy to make inefficient for network file systems - because the control block for each file has to be accessed.  

Of course, we could use alternative structures and it may well be that the tradeoffs have changed sufficiently since the 1970s. Maybe a detailed analysis of the kinds of file traversals and lookups common in a web site would reveal a need for a different design. Similarly, disk drives are different both in scale and operation and flash storage is common. Maybe a hash table would be more efficient. Modern implementations usually involve an in memory hash table used as a cache so that $hash(p)=i$ only if $\Namei(root,p)=i$. This cache design amortizes lookup costs.  

Some of the advantages of the UNIX file-system design are purely semantic. The recursive structure ensures that there are no holes - no paths that terminate at a file that skip over inaccessible directories. This property would require additional work to guarantee in a hash-table implementation if the architects considered it important.  Another set of issues becomes obvious once we consider modifications.
\bibliography{all}
\bibliographystyle{IEEEtran}
\newpage
\begingroup
\parindent 0pt
\parskip 2ex
\def\enotesize{\normalsize}
\endgroup
\end{document}